\documentclass[review]{elsarticle}
\usepackage{lineno,hyperref}
\usepackage{color}
\usepackage[dvipsnames]{xcolor}
\usepackage[normalem]{ulem}
\pdfoutput=1
\modulolinenumbers[5]
\journal{Journal of \LaTeX\ Templates}

\begin{document}
\begin{frontmatter}
\title{Solution of the Anderson chain with two-particle hybridization of localized and itinerant electrons}
\tnotetext[mytitlenote]{Solution of the  Anderson model with two-particle hybridization  of localized and itinerant electrons}
\author{Igor N. Karnaukhov}
\address{G.V. Kurdyumov Institute for Metal Physics, 36 Vernadsky Boulevard, 03142 Kiev, Ukraine}
\fntext[myfootnote]{karnaui@yahoo.com}



\begin{abstract}
A modified Anderson lattice is proposed, whose the Hamiltonian accounts for the two-particle hybridization of localized and itinerant electrons instead of one-particle hybridization which takes into account in the original Anderson model. The 1D version of this model can be solved using the Bethe ansazt. It has been shown that hybridization between itinerant and frozen localized electrons results in an effective on-site interaction between itinerant electrons. The magnitude and sign of this interaction depend on the position of the level relative to the Fermi energy (the energy of this level corresponds to the two-particle state of localized electrons at a site).
It is shown that when the energy of an localized electron in the two-particle state lies above the Fermi energy, the effective on-site interaction between itinerant electrons is attractive. A repulsive interaction between itinerant electrons occurs when this energy lies below the Fermi energy. Thus, two-part hybridization between localized and itinerant electrons can lead to effective attraction between itinerant electrons, which is unique in itself and may underlie the nature of high-temperature superconductivity. 
\end{abstract}
 \begin{keyword}
\texttt  Anderson model \sep Hubbard chain \sep high-temperature superconductivity
\end{keyword}
\end{frontmatter}
\section{Introduction}

It has been 39 years since the discovery of high-temperature superconductivity in cuprates \cite{0}, yet the nature of high-temperature superconductivity remains an unsolved mystery.
The phenomenon of high-temperature superconductivity forces us to think outside the box. To build a new theory explaining unconventional superconductivity, some new approach is required in the theory, where we clearly know what interaction is key in the formation of electron pairs or condensate. Without dwelling on the extensive volume of various explanations of the phenomenon of high-temperature superconductivity, let's simply list the main mechanisms of high-temperature superconductivity proposed: the BCS theory \cite{a1}; the  resonating valence bond theory \cite {a2}; the Fermi surface theory which posits that superconductivity arises from the reconstruction of electronic states, Van Hove singularity theory \cite{a3}; charge density wave theory, superconductivity is associated with charge density waves \cite {a4}. It is critically important to take into account the actual complex structure of high-temperature superconductors in calculations. Simplified models do not fully describe their band structure, yet can provide insight into the nature of electron pairing and the formation of condensate in electron liquid. The superconducting phase, characterized by cohesive electron states (Cooper pairs), is essential for understanding phenomenon of  superconductivity.

The paper proposes a modification of the lattice Anderson model. It is believed that only localized electrons in a two-particle state hybridize with itinerant electrons, as the energy of a localized electron lies outside the conduction band. In this case, there is no one-particle hybridization of itinerant and localized electrons; instead, two-particle hybridization of electrons should be taken into account. It should be noted that for the first time, two-particle hybridization between itinerant and localized electrons has been studied. Within this model, the effective on-site interaction between itinerant electrons is calculated. If the two-particle state of $d$-electrons is empty, the two-particle interaction between the itinerant electrons is attractive, leading to the formation of bound electron pairs.

 \section{The model Hamiltonian}

The Hamiltonian of the modified Anderson chain model ${\cal H}$ has the following form
\begin{eqnarray}
&&{\cal H}=- \sum_{\sigma}\sum_j (c^\dagger_{j,\sigma}c_{j+1,\sigma} +c^\dagger_{j+1,\sigma}c_{j,\sigma})+\varepsilon_g \sum_{\sigma}\sum_j n_{j,\sigma} +\nonumber \\&&
v \sum_j (c^\dagger_{j,\uparrow}c^\dagger_{j,\downarrow}d_{j,\uparrow}d_{j,\downarrow}+
d^\dagger_{j,\downarrow}d^\dagger_{j,\uparrow}c_{j,\downarrow}c_{j,\uparrow})+ U\sum_j n_{j,\uparrow}n_{j,\downarrow},
\label{eq:1}
\end{eqnarray}
where $c^\dagger_{j,\sigma},c_{j,\sigma}$ and  $d^\dagger_{j,\sigma},d_{j,\sigma}$ $(\sigma=\uparrow,\downarrow)$ are the Fermi operators of the conduction and localized electrons defined on the site $j$, $\varepsilon_g$ is the energy of an $d$-electron, $U$ is the  strength of the on-site Hubbard interaction for $d$-eelctrons, and $n_{j,\sigma}=d^\dagger_{j,\sigma}d_{j,\sigma}$ and $m_{j,\sigma}=c^\dagger_{j,\sigma}c_{j,\sigma}$ are the density operators for localized and conduction electrons. Parameter $v$  determines two-particle on-site hybridization  between localized and conduction electrons. The Hamiltonian (\ref{eq:1}) conserves the total  number of particles $N_e=\sum_{\sigma}\sum_j (n_{j,\sigma}+m_{j,\sigma})$.

In the single impurity Anderson model, $d$-electrons can be in four states: $\vert 0\rangle$, $ d^\dagger_{j\uparrow}\vert 0\rangle$, $ d^\dagger_{j\downarrow}\vert 0\rangle$ and $d^\dagger_{j\uparrow}d^\dagger_{j\downarrow}\vert 0\rangle$ \cite{TW,IK1}. The energies corresponding to these states lie  in the conduction band, so the conduction electrons hybridise with the $d-$ electrons in all states. With strong local Hubbard interaction (i.e., in the limit $U\to\infty$), the energy of an electron pair located at the same point $ d^\dagger_{j\uparrow}d^\dagger_{j\downarrow}\vert 0\rangle$ is above the conduction band, so the conduction electrons  hybridise with three states of $d$-electrons whose energies lie in the conduction band.
In the proposed model, the energy of states with one $d$-electron lies outside the conduction band, so conduction electrons do not hybridise with $d$-electrons in these states.  The model Hamiltonian takes into account the two-particle hybridization of conduction electrons with two $d$-electrons located at a site (see Fig. 1). In the lattice model (\ref{eq:1}),  $d$-electrons are frozen, and  conduction electrons move in an effective potential formed by the on-site interaction between conduction and localized electrons.
\begin{figure}[tp]
      \centering{\leavevmode}
\begin{minipage}[h]{.49\linewidth}
\center{
q\includegraphics[width=\linewidth]{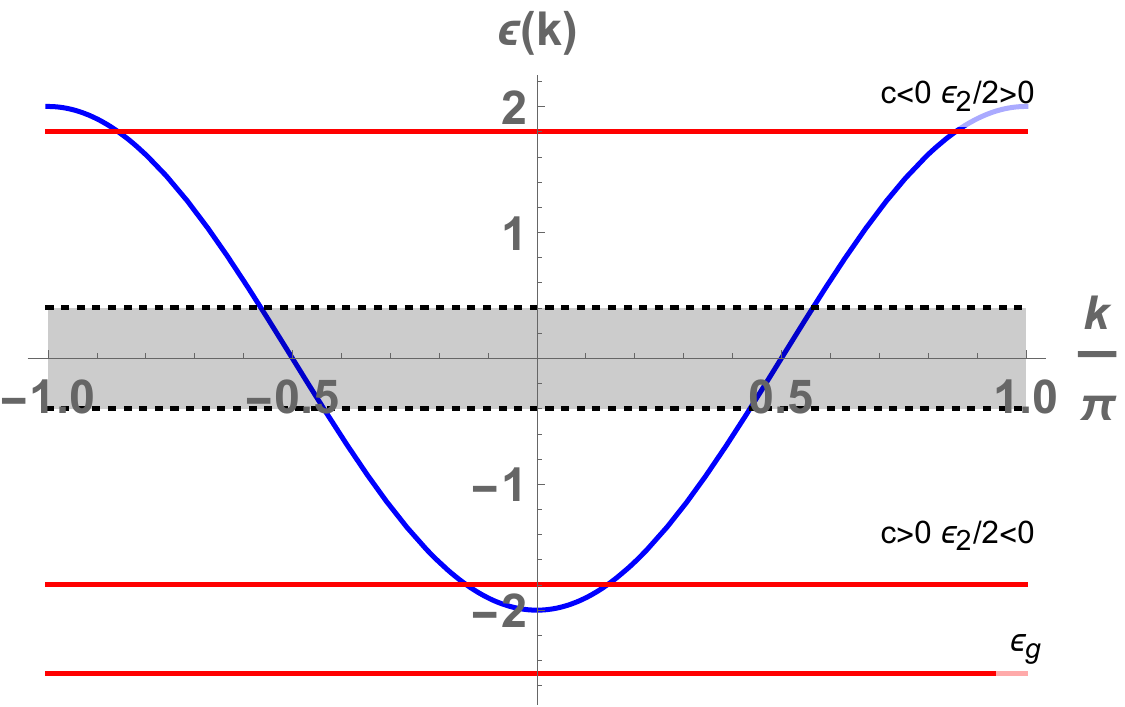} 
                  }
    \end{minipage}
\caption{The one-particle energies of conduction  (blue line) and localized electrons (red lines) as a function of the wave vector, $\varepsilon_2=2\varepsilon_g +U$. The energy region near the Fermi energy (equal to zero), where the effective on-site interaction between conduction electrons (equal to $2c$) does not depend on the wave vectors of scattered electrons, is shaded.
}
\label{fig:1}
\end{figure}

\section{Solution of the problem}

We start with one-particle problem, the wave function corresponds to the following electron states:  $\psi(j,\sigma) c^\dagger_{j,\sigma}$, $\varphi(j,\sigma) d^\dagger_{j,\sigma}$. The amplitudes $\psi$, $\varphi$ satisfy the following equations:
\begin{eqnarray}
&& \epsilon  \psi(j,\sigma) =- \sum_{\textbf{1}}\psi(j+\textbf{1},\sigma), \nonumber\\
&& (\epsilon -\varepsilon_g)\varphi(j,\sigma)=0, 
\label{eq:2}
\end{eqnarray}
where $\epsilon$ is the energy of a given state, the sum is over the nearest lattice sites.

Eqs~(\ref{eq:2}) determine the energy of a conduction electron, $\epsilon(k)=-2\cos k$, $k$ is the momentum of an electron, the energy of a localized electron is equal to $\epsilon =\varepsilon_g$. One-particle state of $d$-electrons is frozen, it does not hybridize with conduction electron.

The eigenvector of the Hamiltonian (\ref{eq:1}) for two particles, which defines the singlet state of electrons, has the following form
\begin{eqnarray}
&& \vert \Psi>=\sum_{j_1,j_2} \psi(j_1,\uparrow;j_2,\downarrow)c^\dagger_{j_1,\uparrow}c^\dagger_{j_2,\downarrow}\vert0>+ \sum_{j_1,j_2} \varphi(j_1,\uparrow;j_2,\downarrow)d^\dagger_{j_1,\uparrow}d^\dagger_{j2,\downarrow}\vert 0>,
\label{eq:3}
\end{eqnarray}
where the amplitudes $\psi(j_1,\uparrow;j_2,\downarrow)$ and $\varphi(j_1,\uparrow;j_2,\downarrow)$ satisfy the Schr\"{o}dinger equation for $j_1\neq j_2$
\begin{eqnarray}
&&
\epsilon \psi (j_1,\uparrow;j_2,\downarrow)+
\sum_{\textbf{1}}[\psi(j_1+\textbf{1},\uparrow;j_2,\downarrow) + \psi(j_1,\uparrow;j_2+\textbf{1},\downarrow)]=0,
\nonumber\\&&
 (\epsilon -2\varepsilon_g)\varphi (j_1,\uparrow;j_2,\downarrow)=0.
 \label{eq:4}
 \end{eqnarray} 
Two-particle state of the conduction electrons has energy $\epsilon= \varepsilon(k_1)+\varepsilon (k_2)$, here  $k_1$ and $k_2$ are the momenta of the conduction electrons, located at the lattice sites $j_1$ and $j_2$. We believe  that the localized electrons in the two-particle state are hybridized with the conduction electrons, the energy $\varepsilon_g$ lies below the conduction band (see  in Fig.1 ). We write the equations for the amplitudes of the wave function $\psi(j_1,\uparrow;j_2,\downarrow)$ and $\varphi(j_1,\uparrow;j_2,\downarrow)$  at $j_1= j_2=j$
\begin{eqnarray}
&&
[\epsilon (k_1)+ \epsilon (k_2)]\psi (j,\uparrow;j,\downarrow)+
\sum_{\textbf{1}}[\psi(j+\textbf{1},\uparrow;j,\downarrow) +
 \psi(j,\uparrow;j+\textbf{1},\downarrow)]=v\varphi(j,\uparrow;j,\downarrow),\nonumber \\&&
   [\epsilon (k_1)+ \epsilon (k_2)-2\varepsilon_g-U]\varphi (j,\uparrow;j,\downarrow)=
v \psi (j,\uparrow;j,\downarrow).
 \label{eq:5}
 \end{eqnarray}

The solution for the two-particle wave function is determined by the Bethe ansatz in a traditional form 
$$\psi (j_1,\sigma_1;j_2,\sigma_2)=A_{\sigma_1\sigma_2}(12)\exp(ik_1j_1+ik_2j_2)-A_{\sigma_1\sigma_2}(21)\exp(ik_1j_2+ik_2j_1)$$ for $j_1<j_2$ and
$$\psi (j_1,\sigma_1;j_2,\sigma_2)=A_{\sigma_2\sigma_1}(21)\exp(ik_1j_1+ik_2j_2)-A_{\sigma_2\sigma_1}(12)\exp(ik_1j_2+ik_2j_1)$$ for $j_1>j_2$, where the solution for the amplitudes $A_{\sigma_1\sigma_2}(12)$, $A_{\sigma_1\sigma_2}(21)$,
$A_{\sigma_2\sigma_1}(12)$ and $A_{\sigma_2\sigma_1}(21)$  
determines a two-particle scattering matrix ${\cal S}_{12}$ of the conduction electrons. 
The Bethe function $\psi(j_1,\sigma_1;j_2,\sigma_2)$ is continuous at $j_1=j_2$, because the amplitudes satisfy the equation $A_{\sigma_1\sigma_2}(12)-A_{\sigma_1\sigma_2}(21)=A_{\sigma_2\sigma_1}(21)-A_{\sigma_2\sigma_1}(12)$. According to  (\ref{eq:1})  the triplet states of electrons do not scatter. 

The solution for the Bethe function follows from Eq.~(\ref{eq:5}), and the two-particle scattering matrix is defined as
\begin{eqnarray}
{\cal S}_{12}=\frac{\sin k_1-\sin k_2+ i c (k_1,k_2){\cal P}_{12} }{\sin k_1-\sin k_2 +i c (k_1,k_2)},
\label{eq:6}
\end{eqnarray}
where ${\cal P}_{12}$ is the spin-permutation operator,  $c(k_1,k_2)=\frac{1}{2}\frac{v^2}{\epsilon (k_1)+\epsilon (k_2)-2\varepsilon_g -U}$ defines an effective on-site interaction between conduction electrons.
The two-particle scattering matrix (\ref{eq:6}) describes the scattering of two conduction electrons in singlet state, as it takes place in the Hubbard chain  \cite{LW}, an effective on-site interaction between conduction electrons is equal to $2c(k_1,k_2)$. In contrast to traditional exactly solvable models  \cite{TW,IK1,LW,IK2,HM}, in which the scattering potential does not depend on the wave vectors of electrons, $c(k_1,k_2)$ is a function of the momenta of scattered electrons. As we noted above, in the conduction band  there is only one level (see in Fig.1), that corresponds to the energy of two d-electrons $\epsilon_2=2\varepsilon_g+U$. 

We will consider the states of conduction electrons with energies close to the Fermi energy, while the energy of a $d$-electron $\epsilon_2/2$ lies far from the Fermi energy, thus $\vert \epsilon_2/2 \vert>>\vert\epsilon (k)\vert$ for $\vert k-k_F \vert <\delta$, where $k_F$ is the Fermi momentum, $\delta$ determines the range of electron momentum changes (see Fig. 1 for an explanation of the above statement). In this approximation, the effective on-site interaction does not depend on the wave vectors of the scattered electrons and $c=-\frac{1}{2}\frac{v^2}{2\varepsilon_g +U}$. The sign of the effective on-site interaction depends on the position of energy $\epsilon_2/2$ a relative to the Fermi energy or the filling of two-particle  $d$-state. So when this level lies above the Fermi energy, an effective attraction between conduction electrons is realised, when this level lies below the Fermi energy, electrons are repulsed by this effective interaction (see in Fig. 1). Unlike Anderson's chain, in this model the $d$-states are frozen, and the model has an exact solution for $c$-interaction, where $c$ is a constant. In this case, we will study the behaviour of electron liquid at an arbitrary interaction value. Provided that the value of $c$ does not depend on the wave vectors of the scattered electrons, one can use the exact solution of the Hubbard chain obtained by Lieb and Wu \cite{LW}. 

The problem reduces to solving a set of ${N_e}+M$ coupled algebraic equations for the $N_e$
quasi-momenta $k_j$ and $M$ rapidities $\lambda_\alpha$, the Bethe equations for these unknown read
\begin{eqnarray}
&&\exp( ik_j N)=\prod_{\alpha=1}^{M}\frac{\lambda_\alpha-\sin k_j -i \frac{c}{2}}{\lambda_\alpha-\sin k_j +i \frac{c}{2}}, \nonumber \\
&&\prod_{j=1}^{N_e}\frac{\lambda_\alpha-\sin k_j -i \frac{c}{2}}{\lambda_\alpha-\sin k_j +i \frac{c}{2}}=-
\prod_{\beta=1}^{M}\frac{\lambda_\alpha-\lambda_\beta -i c}{\lambda_\alpha-\lambda_\beta+i c}, 
\label{eq:7}
\end{eqnarray}
where $N$ is the total number of lattice sites, $M$ is the total number of electrons with spin upper.

In the thermodynamic limit ($N\to \infty$, with $\rho =N_e /N$ and $h=M/N$ being fixed) the structure of the solutions of
the Bethe  equations (\ref{eq:7}) for the ground-state includes real quasi-momenta for $c>0$ and complex quasi-momenta for $c<0$.

For repulsive effective interaction at half filling the Mott insulator state is realized. According to \cite{LW} the charge gap $\Delta$  in the spectrum of conduction electrons is equal to 
 \begin{eqnarray}
&&\Delta= 2c-4 +8 \int_{0}^{\infty}d \omega\frac{J_1(\omega)}{\omega[1+\exp (\omega c)]}, 
\label{eq:8}
\end{eqnarray}
where $J_1(\omega)$ is the Bessel function of order one.

This gap does not correspond to the Kondo insulator state, since $d$-electrons do not form a local moment in the singlet state, and conduction electrons do not scatter on local magnetic moments. The gap $\Delta$  has the Mott nature, spin excitations are gapless.

In the attractive interaction $c<0$ the ground state is filled with bound states, they are determined by the following complex solutions of the Bethe equations (\ref{eq:7}) in the ground-state:
\begin{eqnarray}
&& \sin( k^\pm_j )= \lambda_j \pm i\frac{c}{2} +0(\exp(-\eta N)), Im \lambda_j=0, \eta>0. 
\label{eq:9}
\end{eqnarray}
The spin excitations has a gap $\Delta$, the charge excitations (holon and antiholon) are gapless.

The main feature to note is that superconducting correlations dominate in this case, their an asymptotic behavior is defined as
\begin{eqnarray}
&& <c^\dagger_{j,\uparrow}c^\dagger_{j,\downarrow}c_{1,\uparrow}c_{1,\downarrow}>\sim \frac{1}{j^\nu}, j\to \infty,
\label{eq:10}
\end{eqnarray}
where $\frac{1}{2}\leq \nu \leq 1$, the value of $\nu$ approaches $\frac{1}{2}$ in the low density limit, it goes to 1 for half filled band. 
For comparision the correlation function $<c^\dagger_{j,\sigma}c_{1,\sigma}>$ decays exponentially, and superconducting correlation function decays slower then density-density correlation function $<m_{j,\sigma}m_{1,\sigma}>$ \cite{HM}
\begin{eqnarray}
&&<m_{j,\sigma}m_{1,\sigma}>-<m_{1,\sigma}>^2 \sim  \frac{\cos(2\pi j \rho)}{j^{1/\nu}}, j\to \infty.
\label{eq:11}
\end{eqnarray}

The mechanism of high-temperature superconductivity formation should be simple enough for calculation and understanding.
Phenomenon of Cooper pairing is a good example of what was said above.
A natural question arises as to why the sign of effective interaction depends on the position of the $d$-electron level relative to the Fermi energy. When the two-particle $d$-level lies above the Fermi energy  ($d$-state is empty $n_{j,\uparrow}=n_{j,\downarrow}=0$ at $v=0$), the conduction electrons at the site hybridize with the $d$-electrons, so their effective charge becomes smaller. As a result, the effective repulsion between them decreases. For $\varepsilon_2 <0$, the $d$-level is filled ($d$-state is full-filled  $n_{j,\uparrow}=n_{j,\downarrow}=1$ at $v=0$), and this charge compensation for conduction electrons does not occur. Effective charge of conduction electrons increases, effective on-site interaction between electrons is repulsive.

From the solutions of the Bette equations (\ref{eq:7}), it follows that in the ground state, under attractive effective interactions between conduction electrons, bound two-particle states of conduction electrons are formed, superconducting correlations dominate in the electron liquid.  In two- and three-dimensional systems, the hybridization mechanism of electron pairing should lead to a superconducting state of the electron liquid, since we are talking about the effective on-site attraction of conduction electrons. The electron nature of the interaction allows us to speak of a high superconducting temperature, i.e. high-temperature superconductivity.

\section{Conclusion}
We have calculated the ground state of  the Anderson chain, in which itinerant and localized electrons interact  via two-particle hybridization. When the energy of one electron in two $d$-electron state, is far from the Fermi energy, the 1D model has solution using the Bethe ansatz. The effective interaction between conduction electrons can be either repulsive (if the two-electron $d$-level is filled) or attractive (if it is empty). The solution obtained  shows that the behaviour of the electron liquid in the 1D model has superconducting properties, with superconducting fluctuations dominating, singlet pairs of conduction electrons forming in the ground state.

The proposed mechanism for forming effective interaction between conduction electrons through their two-particle hybridization with localized electrons does not depend on the dimension of the system. This suggests that this is precisely the nature of electron-electron attraction of itinerant electrons in real high-temperature superconductors.
Our analytical results for the 1D modified Anderson chain provide new insights into the behavior of the electron liquid of many-band electron systems.

\subsection*{Author contributions statement}
I.K. is the author of the manuscript
\subsection*{Additional information}
The author declares no competing financial interests. 
\subsection*{Availability of Data and Materials}
All data generated or analysed during this study are included in this published article.

\end{document}